\input{aipcheck}

\documentclass[
    ,final            
  ]
  {aipproc}

\layoutstyle{6x9}

\def\siml{{\ \lower-1.2pt\vbox{\hbox{\rlap{$<$}\lower6pt\vbox{\hbox{$\sim$}}}}\ }}
\def\simg{{\ \lower-1.2pt\vbox{\hbox{\rlap{$>$}\lower6pt\vbox{\hbox{$\sim$}}}}\ }}

\def \als {\alpha_{\mathrm{s}}}

\def \m2   {\mu^{2 \epsilon}}
\def\lQ{\Lambda_{\rm QCD}}
\def\als{\alpha_{\rm s}} 
\def\bea{\begin{eqnarray}}
\def\eea{\end{eqnarray}}
\def\be{\begin{equation}}
\def\ee{\end{equation}}

\begin{document}

\title{Heavy Quarkonium at zero and finite temperature: an effective field theory perspective}


\author{Nora Brambilla}{
  address={Physik-Department, Technische Universit\"at M\"unchen,
James-Franck-Str. 1, 85748 Garching, Germany}
}

\begin{abstract}
I discuss quarkonium physics at zero and finite temperature in 
the framework of nonrelativistic effective field theories.
\end{abstract}

\maketitle


\section{Quarkonium Physics and EFTs}

Effective field theories for the description of quarkonium processes  have been newly developed and are being 
developed and  provide a  unifying description
as well as a solid and versatile tool giving well definite predictions
\cite{Brambilla:2004jw,Brambilla:2004wf,qwg2}.
They rely on one  hand on  high order perturbative calculations and on 
the other hand on lattice simulations, the recent  progress in both fields 
having added a lot to the theory reach.
Heavy quarkonium is a multiscale system and as such is on one hand 
particularly suitable to be treated in an effective field theory framework. 
On the  other hand the existence of many scales  in quarkonium makes it a unique system to 
study complex environments. Quarkonium probes all the regimes of QCD, from the 
high energy region, where an expansion in the the coupling constant is 
possible, to the low energy region, where nonperturbative effects dominate. It 
probes also the intermediate region between the two regimes.
In particular for quarkonium system with a very small radius the 
interaction turns out to be purely perturbative 
while for system with a large radius 
with respect to the confinement scale the interaction turns out to be 
nonperturbative.
In the complex environment of heavy ion collisions
quarkonium suppression constitutes  a unique probe  of deconfinement and 
quark gluon plasma formation \cite{Matsui:1986dk}. 
The different radius of the different 
quarkonia states induces the phenomenon of sequential suppression, 
allowing to use quarkonium as  a kind of thermometer for the measurement of the temperature of the formed 
medium. On similar ground quarkonium may constitute  a  special probe
to be used in the study  of a  nuclear medium.
The large mass, the clean and known decays mode make quarkonium an ideal
probe of new physics in some well defined  window of parameters of physics beyond the 
Standard Model \cite{Brambilla:2004wf,qwg2}.

The modern approach to quarkonium physics consists in taking  
advantage of the hierarchy of non-relativistic energy scales in the system 
by constructing a suitable hierarchy of effective field theories (EFTs)~\cite{Brambilla:2004jw}.

The energy scales are:  the heavy-quark mass (hard scale), 
$m$, the typical momentum transfer (soft scale), 
$p \sim mv$, whose inverse sets the typical distance, 
$r$, between the heavy quark and the antiquark, 
and the typical kinetic energy (ultrasoft scale), 
$E \sim mv^2$, whose inverse sets the typical time scale of the bound state. 
The heavy-quark bound-state velocity $v$ is a small quantity $v\ll 1$
($v^2 \sim 0.1$ for $b\bar{b}$, $v^2 \sim 0.3$ for $c\bar{c}$,  
$v^2 \sim 0.01$ for $t\bar{t}$), the mass is a large quantity $m\gg \lQ$,
$\als(m) \ll 1$.
For energy scales close to $\lQ$, perturbation theory breaks down  
and one has to rely on nonperturbative 
methods. Regardless of this, the nonrelativistic hierarchy of scales: 
$m \gg p \sim 1/r \sim mv  \gg E \sim m v^2$
also persists below the $\lQ$ threshold.
 While the hard scale is always  larger than 
$\lQ$, different  situations may arise for the other two scales 
depending on the considered quarkonium system.
The  soft scale, proportional to the  inverse typical radius $r$,
may be a perturbative ($\gg \lQ$) or a nonperturbative scale ($\sim \lQ$) depending on the physical system. 
The first case is likely to happen only for the lowest charmonium and 
bottomonium states. We do not have direct
information on the radius of the quarkonia systems, and thus the
attribution of  some of the lowest bottomonia and charmonia states 
to the perturbative or the nonperturbative soft regime is at the
moment still ambiguous.
The ultrasoft scale  may still be perturbative only in 
the case of  $t\bar{t}$ threshold states.
All quarkonium scales get entangled in a typical amplitude involving a quarkonium
observable. In particular, quarkonium annihilation 
and production happen at the scale $m$, quarkonium binding happens at the scale
$mv$, which is the typical momentum exchanged inside the bound state, while 
very low-energy gluons and light quarks (also called ultrasoft degrees of freedom)  
live long enough that a bound state has time to form and, therefore, are sensitive to the 
scale $mv^2$. Ultrasoft gluons are responsible for phenomena 
like the Lamb shift in QCD.

A hierarchy of EFTs may be constructed by systematically integrating out 
modes associated to  high energy scales not relevant for quarkonium.
Such integration  is made  in a matching procedure enforcing 
the  equivalence between QCD and the EFT at a given 
order of the expansion in $v$.
The EFT  realizes a factorization at the Lagrangian level between 
the high energy contributions, encoded into the  matching coefficients, and 
 the low energy contributions, carried by the dynamical degrees of freedom.
Poincar\'e symmetry remains  intact in a nonlinear realization at the level of the NR EFT
and imposes exact relations among the 
matching coefficients  \cite{Brambilla:2003nt}.

At the scale $m$ the suitable EFT is NRQCD  \cite{Caswell:1985ui}, which 
follows from QCD by integrating out the scale $m$. As a consequence, 
the effective Lagrangian is organized as an expansion in $1/m$  and $\als(m)$.
The field of quarkonium production has seen terrific progress in the last 
few years both in theory and in experiments, for a review see  \cite{Brambilla:2004wf,qwg2}.

For what concerns decays,  recently, substantial progress has been made in the evaluation of the factorization formula at order $v^7$ \cite{Brambilla:2006ph}, in the lattice evaluation of the NRQCD matrix elements 
\cite{Bodwin:2005gg} and in the data of many hadronic 
and electromagnetic decays \cite{Brambilla:2004wf}. 
The data are clearly sensitive to NLO corrections in the Wilson coefficients 
and presumably also to relativistic corrections. Improved  theory predictability would entail  the lattice 
calculation or data extraction of the NRQCD matrix elements and 
perturbative resummation of large contribution in the NRQCD matching coefficients.
The new data on hadronic transitions and hadronic decays pose interesting 
challenging to the theory.

Lattice NRQCD calculations have undergone a steady development in last few years see \cite{Brambilla:2004wf,qwg2}.


At the scales $mv$ and $mv^2$ 
the suitable EFT is potential 
  NonRelativistic QCD (pNRQCD)   
\cite{Pineda:1997bj,Brambilla:1999xf}, which 
follows from NRQCD by integrating out the scale $mv$. 
As a consequence, the effective Lagrangian is organized as an expansion in $1/m$  and $\als(m)$, 
inherited from NRQCD, and an expansion in $r$
$$
{\cal L}_{\rm pNRQCD}  = \int d^3r\,  
\sum_n \sum_k \frac{c_n(\als(M),\mu)}{M^{n}}  
 V_{n,k}(r,\mu^\prime, \mu) \; r^{k}  \times O_k(\mu^\prime,Mv^2,...) ,
$$
where $O_k$ are the operators of pNRQCD that live at the low-energy scale
 $mv^2$, $\mu$ is the NRQCD factorization scale,
 $\mu^\prime$ is the pNRQCD factorization scale and $V_{n,k}$
are the Wilson coefficients of the EFT that encode the contributions 
from the scale $r$ and are non-analytic in $r$.
Looking at the equations of motion of pNRQCD, we may identify 
$V_{n,0}$ with the $1/m^n$ potentials that enter the Schr\"odinger equation and 
$V_{n,k\neq 0}$ with the couplings of the ultrasoft degrees of freedom, which
provide corrections to the Schr\"odinger equation. 
This EFT  is close to a Schr\"odinger-like description of the bound
state. 
The bulk of the interaction
is carried by potential-like terms $V_{n,0}$, but non-potential interactions
$V_{n,k\neq 0}$,associated with the propagation of low-energy degrees of freedom 
($Q\bar{Q}$ colour singlets, $Q\bar{Q}$  colour  octets and low energy gluons),
are generally present. They  start  to contribute at  NLO 
in the multipole expansion of the gluon fields and are 
typically  related to nonperturbative effects
\cite{Brambilla:1999xf}.

In what follows we will focus on the EFT at the scale $mv$ and $mv^2$.
Then, there are several cases for the physics at hand. 
In the case in which the EFT has been constructed \cite{Brambilla:1999xf,Brambilla:2000gk,Brambilla:2004jw},
 i.e. for states below 
threshold,  the work is currently going in calculating higher 
order perturbative corrections in $v$ and $\als$ for processes of interest,
resumming the logarithms in the ratio of the scales that may be sizeable,
calculating or extracting nonperturbatively low energy correlators and extending the theory with the addictions of electromagnetic effects \cite{Brambilla:2005zw} and  the consideration 
of $QQQ$ and $QQq$ systems \cite{Brambilla:2009cd}.
The issue here is precision physics  and the study of confinement.
Close to threshold the EFT has not yet been constructed and the degrees 
of freedom have still to be identified \cite{Brambilla:2008zz,qwg2}.
At finite temperature the EFT is being constructed and the existing results 
hint at a new physical picture with possible application at heavy ion 
collisions at LHC.

Below we will review new results  focusing on the example of the 
calculation of the  interquark potential at zero and at finite 
temperature in the different dynamical situations.

\section{Quarkonium potential at zero temperature}

For states away from threshold we have a clear effective field 
description called pNRQCD, based on perturbative and lattice computations.
This is nowadays the standard description.

The soft scale $mv$  may be larger or not than the confinement 
scale $\lQ$ depending on the radius of the quarkonium system. 
When $mv^2 \sim \lQ$, we speak about weakly-coupled pNRQCD because 
the soft scale is perturbative and the matching from NRQCD to pNRQCD 
may be performed in perturbation theory. 
When $mv  \sim \lQ$, we speak about  
strongly-coupled pNRQCD because the soft scale 
is nonperturbative and the matching from NRQCD to pNRQCD may not be performed in
perturbation theory.

The potential is a Wilson coefficient  of an EFT. 
In general undergoes renormalization, develops scale dependence and satisfies renormalization 
group equations  which allow to resum large logarithms.

\subsection{weakly-coupled  pNRQCD}
If the quarkonium system is small, the soft scale is perturbative and the 
potentials can be  entirely  calculated in perturbation theory 
\cite{Brambilla:2004jw}.

Since the degrees of freedom that enter the Schr\"odinger description 
are in this case both $Q\bar{Q}$  color singlet and $Q\bar{Q}$ color octets,
both singlet and octet potentials exist.
The static singlet $Q \bar{Q}$ potential is pretty well known.
The three-loop correction to the static potential is now completely
known: the fermionic contributions to the three-loop 
coefficient~\cite{Smirnov:2008pn} first became available, and more
recently the remaining purely gluonic term has been 
obtained~\cite{Anzai:2009tm,Smirnov:2009fh}. 

The first log related to ultrasoft effects arises at three 
loops   \cite{Brambilla:1999qa} . Such logarithm  contribution at N$^3$LO 
and the single logarithm contribution at N$^4$LO may be extracted respectively 
from a one-loop and two-loop  calculation in the EFT and have been calculated 
in \cite{Brambilla:2009bi}.

The perturbative series of the static potential suffers from a renormalon ambiguity 
(i.e. large $\beta_0$  contributions) and from large logarithmic contributions.
The singlet 
static energy,  given by the sum of a constant, the static potential and the ultrasoft 
corrections,
is free from ambiguities of the perturbative series. By resumming the large logs using 
the renormalization  group equations and  comparing it
(at the NNLL) with lattice 
calculations of the static energy one sees 
that the QCD perturbative series converges very nicely 
to and agrees with 
the lattice result in the short range    (up to 0.25~fm) and that no nonperturbative
linear (``stringy'') contribution to the static potential exist \cite{Pineda:2002se,Brambilla:2009bi}.

In particular, the 
recently obtained theoretical expression~\cite{Brambilla:2010pp} 
for the complete QCD static
energy at  NNNLL precision has
been used 
to determine $r_0 \Lambda_{\overline MS}$ by comparison with available lattice
data, where $r_0$ is the lattice scale and $\Lambda_{\overline MS}$
is the QCD scale, obtaining 
$r_0\Lambda_{\overline MS} =0.622^{+0.019}_{-0.015}$  for the zero-flavor case. 
This extraction was previously performed
at the NNLO level (including an estimate at NNNLO) in \cite{Sumino:2005cq}.
The same procedure can be used to obtain a precise evaluation of the
unquenched $r_0 \Lambda_{\overline MS}$ value after short distance unquenched
lattice data for the $Q \overline{Q}$ exist.

The static octet potential is known up to two loops~\cite{Kniehl:2004rk}.
Relativistic corrections to the static singlet potential
have been calculated over the years and are 
summarized in \cite{Brambilla:2004jw}. 

In the case of $QQq$ baryons, the static potential has been determined up to 
NNLO in perturbation theory    \cite{Brambilla:2009cd}   and recently also on the lattice \cite{Yamamoto:2007pf}.
Terms suppressed by powers of  $1/m$  and $r$ in the Lagrangian have been matched 
(mostly) at leading order and used to determine, for instance, the expected 
hyperfine splitting of the ground state of these systems.

In the case of $QQQ$ baryons, the static potential has been determined up to 
NNLO in perturbation theory \cite{Brambilla:2009cd}
and also on the lattice \cite{Takahashi:2000te}. The transition region from 
a Coulomb to a linearly raising potential is characterized in this case also 
by the emergence of a three-body potential apparently parameterized by only one length. 
It has been   shown that in 
perturbation theory a smooth genuine three-body potential shows up at two loops.

\subsection{strongly-coupled  pNRQCD}
If the quarkonium system is large, the soft scale is nonperturbative and the 
potentials cannot be  entirely calculated in perturbation theory 
\cite{Brambilla:2004jw}.
Then the  potential matching coefficients
are obtained in the form of expectation values of gauge-invariant 
Wilson-loop operators. 
In this case, heavy-light meson pairs and heavy hybrids 
develop a mass gap of order $\lQ$ with respect to the energy of the
$Q\overline{Q}$ pair, the second circumstance 
being apparent from lattice simulations.
Thus, away from threshold, 
the quarkonium singlet field $S$ is the only low-energy dynamical 
degree of freedom in the pNRQCD Lagrangian \cite{Brambilla:2000gk,Brambilla:2004jw,Brambilla:2007by} 
(neglecting ultrasoft corrections coming 
from pions and other Goldstone bosons).
The singlet potential $V_S(r)$ can be expanded
in powers of the inverse of the quark mass;
static, $1/m$ and $1/m^2$ terms were calculated long 
ago~\cite{Brambilla:2000gk}.
They involve NRQCD matching coefficients (containing 
the contribution from the hard scale) and low-energy 
nonperturbative parts given in terms
of static Wilson loops and field-strength insertions in the static
Wilson loop
(containing the contribution from the soft scale).
Such expressions correct and generalize previous finding in the Wilson loop approach
\cite{Eichten:1980mw} 
that were typically missing the high energy parts of the potentials, 
encoded into the NRQCD matching coefficients and containing the 
dependence on the logarithms of $m$, and some of the low energy contributions.
Poincar\'e invariance   \cite{Brambilla:2003nt} establishes exact relations between the potentials
 of the type of the  Gromes relation between spin 
dependent and static potentials \cite{Gromes:1984ma}.

In this regime of pNRQCD, we recover the quark potential singlet model. 
However, here the potentials are calculated in QCD by nonperturbative 
matching. Their evaluation requires calculations on the lattice 
or in QCD vacuum models \cite{Brambilla:1999ja}.

Recent progress includes new, precise lattice calculations 
of these potentials 
obtained using the L\"uscher multi-level algorithm \cite{Koma:2007jq}.


The nonperturbative potentials for the $QQQ$ and $QQq$  have been obtained in 
 the second reference of  \cite{Brambilla:2009cd}  and in \cite{Brambilla:1993zw}.

\section{Quarkonium close or above threshold}

In the most interesting region, the region close to threshold where many
new states, conceivably of an exotic nature have been recently discovered, 
no EFT description has yet been constructed nor the appropriate degrees of 
freedom
clearly identified   \cite{Brambilla:2008zz,qwg2}.
 An exception is constituted by the $X(3872)$ that 
displays universal characteristics related to its being  so close to threshold, 
reason for which a
beautiful EFT  description  could be   obtained \cite{Braaten:2009zz}.

The threshold region remains troublesome
also for the lattice,  althought  several excited states calculations have been recently being 
pionereed.

Lattice results about the 
the crosstalk of the static potential with a pair of 
heavy-light mesons in the lattice have recently appeared   \cite{Bali:2009er}
 but further investigations
appear to be necessary.

\section{Quarkonium potential at finite temperature}

The study of quarkonium in media has recently undergone crucial developments.
Large datasets from heavy-ion collisions
have recently  become available at RHIC  displaying new features  related
to the quark gluon plasma formation characteristics like the particular
structure of jet quenching and the  very low viscosity to entropy 
ratio. In particular the  quark gluon plasma looks more like a liquid than a 
plasma and the use of perturbative expansion  appears  to be   justified only at 
temperature bigger than the deconfinement one.

The suppression of quarkonium production in
the hot medium remains one of  the cleanest and most relevant
probe of deconfined matter.

However, the use of quarkonium yields as a hot-medium
diagnostic tool has turned out to be quite challenging
for several reasons.
Quarkonium production has already been found to
be suppressed in proton-nucleus collisions by
cold-nuclear-matter effects, which themselves require
dedicated experimental and theoretical attention. Recombination effects
may  play an additional role and  thus transport properties may become 
relevant to  be considered. Finally,  the heavy quark-antiquark interaction
at finite temperature $T$ has to be obtained from QCD.

For observables only sensitive to gluons and light quarks,
a very successfull EFT called Hard Thermal Loop (HTL)
effective theory has been derived  in the past \cite{Braaten:1989mz}
by integrating out the hardest momenta  propotional to $T$
from the dynamics.  However,   considering also heavy quarkonium in the hot 
QCD medium,  one has to consider in addition to the thermodynamical scales 
in $T$ also the scales of the nonrelativistic bound state and the situation 
becomes more complicate. 

In the last few years years, there has been a remarkable 
progress in constructing EFTs  for quarkonium at finite temperature and 
in rigorously defining 
the quarkonium potential.
In \cite{Laine:2006ns,Laine:2007qy}, the static potential was calculated 
in the regime $T \gg 1/r \simg m_D$, where $m_D$ is the Debye mass and 
$r$ the quark-antiquark 
distance, by performing an analytical continuation of the 
Euclidean Wilson loop
to real time. The calculation was done in 
the weak-coupling resummed perturbation
theory. The imaginary part of the gluon self energy gives an imaginary part to
the static potential and hence a thermal width to
the quark-antiquark bound state. In the same framework, the dilepton
production rate for charmonium and bottomonium was calculated in 
\cite{Laine:2007gj,Burnier:2007qm}.
In \cite{Beraudo:2007ky}, static particles in real-time formalism were 
considered  and the potential for distances $1/r \sim m_D$ was derived 
for a hot QED plasma. 
The real part of the static potential was found to agree with the
singlet free energy and the damping factor with the one found in 
\cite{Laine:2006ns}.
In \cite{Escobedo:2008sy}, a study of bound states in a hot QED 
plasma was performed in a non-relativistic EFT framework. 
In particular, the hydrogen atom was studied for temperatures ranging from 
$T\ll m\alpha^2$ to $T\sim m$, where the imaginary part of the
potential becomes larger than the real part and the hydrogen ceases to exist. 
The same study has been extended to muonic hydrogen in 
 \cite{Escobedo:2010tu}, providing a method to estimate  the effects of a 
finite charm quark mass on the dissociation temperature of bottomonium.

An EFT framework in real time and weak coupling for quarkonium at finite
temperature was developed in \cite{Brambilla:2008cx}   working in real time and 
in the regime 
of small coupling $g$, so that $g T \ll T$  and  $v \sim \als$, which is expected to be valid 
for tightly bound states: $\Upsilon(1S)$, $J/\psi$, ...~.

 Quarkonium in a medium is characterized by different energy and momentum scales;  there are the scales of the non-relativistic bound state  that we have  discussed at the beginning,
and there are the  
thermodynamical scales: the temperature $T$, the inverse of the screening 
length of the chromoelectric interactions, i.e. the Debye mass $m_D$ and lower scales, which 
we will neglect in the following.

If these  scales are hierarchically ordered,  then we may expand physical observables in the 
ratio of such scales. If we separate explicitly the contributions from the different scales
at the Lagrangian level this amounts to substituting QCD with a hierarchy of EFTs, which are equivalent 
to QCD order by order in the expansion parameters.
As it has been described in the previous sections
 at zero temperature the EFTs
that follow from QCD by integrating out the scales $m$ and $mv$ are called respectively 
Non-relativistic QCD (NRQCD) and potential NRQCD (pNRQCD).
We assume that the temperature is high enough that $T \gg gT \sim m_D$ holds 
but also that it is low enough for $T \ll m$ and $1/r \sim mv \simg m_D$ to be satisfied, 
because for higher temperature  the bound state ceases to exist.
Under these conditions some possibilities are in order. If $T$ is the next relevant scale after 
$m$, then integrating out $T$ from NRQCD leads to an EFT that we may name NRQCD$_{\rm HTL}$, because 
it contains the hard thermal loop (HTL) Lagrangian
 \cite{Braaten:1989mz}.
Subsequently integrating out 
the scale $mv$ from NRQCD$_{\rm HTL}$ leads to a thermal version of pNRQCD that we may call 
pNRQCD$_{\rm HTL}$. If the next relevant scale after $m$ is $mv$, 
then integrating out $mv$ from NRQCD leads to pNRQCD. If the temperature is larger than $mv^2$, 
then the temperature may be integrated out from pNRQCD leading to a new version of pNRQCD$_{\rm HTL}$  \cite{Vairo:2009ih}.
 Note that, as long as the temperature is smaller than the scale being 
integrated out, the matching leading to the EFT may be performed putting the
 temperature to zero.

The derived potential $V$  describes the real-time evolution of a quarkonium state
in a thermal medium. At leading order, the evolution is governed 
by a Schr\"o\-din\-ger equation. In an EFT framework,
the potential follows naturally from integrating out all  contributions coming from modes
with energy and momentum larger than the binding energy.
For $T < V$ the potential is simply the Coulomb potential. Thermal corrections 
affect the energy and induce a thermal width to the quarkonium state; these 
may be relevant to describe the in medium modifications of quarkonium at low temperatures.
For $T >V$ the potential gets thermal contributions, which are both real and imaginary.

General findings in this picture  are:
\begin{itemize}
\item{}  The thermal part of the potential has a real and  an imaginary part. 
The imaginary part of the potential smears out the bound state 
peaks of the quarkonium spectral function, leading to their dissolution prior
to the onset of Debye screening in 
the real part of the potential (see, e.g. the 
discussion in \cite{Laine:2008cf}).
So quarkonium dissociation appears to  be a consequence of the 
appearance of a thermal decay width rather than being due to the color screening of
the real part of the potential; this follows from the observation that the
thermal decay width becomes as large as the binding energy at a temperature 
at which color screening may not yet have set in.
\item{} Two mechanisms contribute to the thermal decay width: the imaginary part of the gluon self energy  induced by the Landau-damping phenomenon (existing also in QED)  
\cite{Laine:2006ns}  and the quark-antiquark color singlet to color 
octet thermal break up (a new effect, specific of QCD)  \cite{Brambilla:2010xn}.
 Parametrically, the first mechanism dominates for temperatures 
such that the Debye mass $m_D$ is larger than the binding energy, while the latter 
dominates for temperatures such that $m_D$  is smaller than the binding energy.
\item{} The obtained singlet thermal potential, $V$, is neither the color-singlet quark-antiquark free energy  nor the internal energy. It has an imaginary part and may contain divergences
that eventually cancel in physical observables \cite{Brambilla:2010xn}.
\item{} Temperature effects can be other than screening, typically they  may appear as
power law corrections or a logarithmic dependence \cite{Brambilla:2010xn,Escobedo:2008sy}.
\item{} The dissociation temperature goes parametrically as  $\pi T_{\rm melting} \sim m g^{4\over 3}$ \cite{Escobedo:2008sy,Laine:2008cf}.
\end{itemize}

The EFT  provides a clear definition of the potential and a  coherent and systematical setup
to calculate masses and widths of quarkonium at finite temperature.
 In \cite{Brambilla:2010vq} 
heavy quarkonium energy levels and decay widths in a quark-gluon plasma, 
below the melting temperature at a
temperature $T$ and screening mass $m_D$ satisfying the hierarchy  
$m \als  \gg \pi T  \gg m \als^2 \gg m_D$, have been calculated  at order $m \als^5$.
This  situation is relevant for bottomonium $1S$   states ($\Upsilon(1S)$, $\eta_b$) at the LHC. 
It has been found \cite{Brambilla:2010vq} 
 that: at leading order the quarkonium masses increase quadratically with $T$ 
which in turn implies the same functional increase in the energy of the dileptons produced in the electromagnetic decays;  a thermal correction proportial to $T^2$ appears in the electromagnetic quarkonium decay  rates;  at leading order a decay width linear with the temperature is developed which implies a tendency to dissolve by decaying to the continuum of the colour-octet states. 

In \cite{Brambilla:2010xn,Burnier:2009bk} the Polyakov loop  and the correlator of two Polyakov loops 
at finite temperature  has ben calculated at next-to-next-to-leading order in the weak coupling 
regime and at quark-antiquark distances shorter than the inverse of the temperature and for 
Debye mass larger than the Coulomb potential.  
The calculation has been performed also the in EFT framework 
\cite{Brambilla:2010xn} and a relation between the Polyakov loop correlator and the  singlet and octet  quark-antiquark correlator has been established in this setup \cite{Brambilla:2008cx}.

First attempts to generalize this new picture to the nonperturbative 
regime have been undertaken in \cite{Rothkopf:2009pk}.



\begin{theacknowledgments}
We acknowledge financial support from the DFG cluster of excellence ``Origin and structure of the universe''
({http://www.universe-cluster.de}).
\end{theacknowledgments}



\bibliographystyle{aipproc}   


\IfFileExists{\jobname.bbl}{}
 {\typeout{}
  \typeout{******************************************}
  \typeout{** Please run "bibtex \jobname" to optain}
  \typeout{** the bibliography and then re-run LaTeX}
  \typeout{** twice to fix the references!}
  \typeout{******************************************}
  \typeout{}
 }

\end{document}